\documentclass[journal=jacsat,manuscript=article]{achemso}

\usepackage{comment}
\usepackage{chemformula} 
\usepackage[T1]{fontenc} 
\usepackage{graphicx} 
\usepackage{subcaption} 
\usepackage[percent]{overpic} 
\usepackage{bm} 

\author{Ying You}
\affiliation[Stony Brook University]
{Department of Chemistry, Stony Brook University, Stony Brook, NY 11790}
\alsoaffiliation[IACS]
{Institute for Advanced Computational Science, Stony Brook University, Stony Brook, NY 11790}
\author{James K. McCusker}
\affiliation[MSU]
{Department of Chemistry, Michigan State University, East Lansing, MI 48824}
\author{Arshad Mehmood}
\affiliation[IACS]
{Institute for Advanced Computational Science, Stony Brook University, Stony Brook, NY 11790}
\alsoaffiliation[DoIT]
{Division of Information Technology, Stony Brook University, Stony Brook, NY 11790}
\author{Benjamin G. Levine}
\affiliation[Stony Brook University]
{Department of Chemistry, Stony Brook University, Stony Brook, NY 11790}
\alsoaffiliation[IACS]
{Institute for Advanced Computational Science, Stony Brook University, Stony Brook, NY 11790}
\email{ben.levine@stonybrook.edu}

\title[Cr(III) coordination complex dynamics]
  {Vibrations Drive Ultrafast Intersystem Crossing of a Photoexcited Cr(III) Complex}

\abbreviations{}

\begin{document}

\begin{tocentry}

\includegraphics[width=\linewidth]{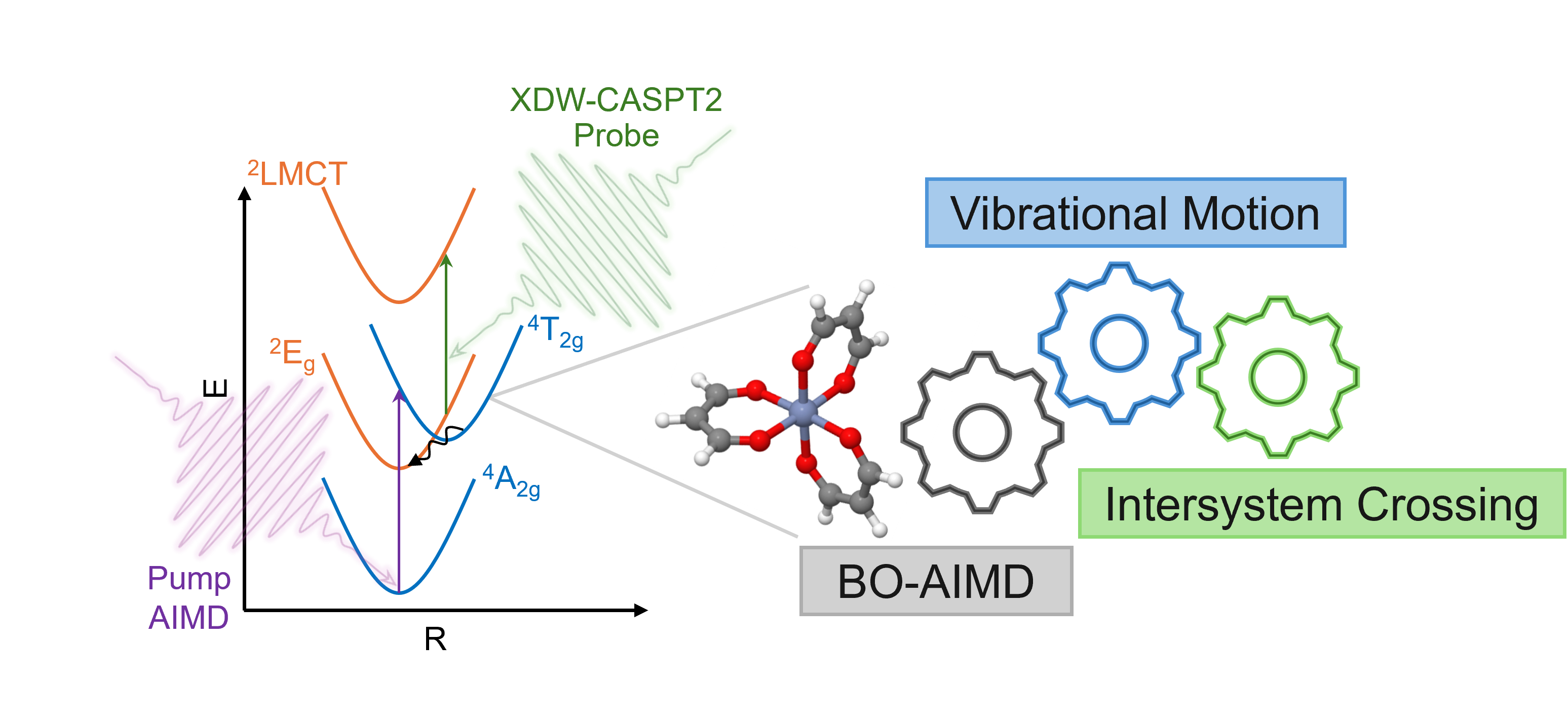}

\end{tocentry}

\begin{abstract}
  
The Cr(III) coordination complex serves as an archetypical 3d transition metal system for probing ultrafast excited-state dynamics with spin conversion due to its intrinsic intersystem crossing (ISC) pathway, \textsuperscript{4}T\textsubscript{2g} $\rightarrow$ \textsuperscript{2}E\textsubscript{g}, upon photoexcitation. Here we conduct \textit{ab initio} molecular dynamics simulations in the \textsuperscript{4}T\textsubscript{2g} state of a model Cr(III) coordination complex, followed by analyses of multireference electronic structure properties. Across 50 trajectories, the compound retains a persistent Jahn--Teller distortion in the excited state, while exhibiting prominent symmetric metal-ligand bond stretching vibrations with frequencies of 219\,cm$^{-1}$ and 465\,cm$^{-1}$. State-averaged complete active space self-consistent field (SA-CASSCF) calculations obtain two corresponding normal modes at 225\,cm$^{-1}$ and 487\,cm$^{-1}$ with symmetric stretching character. The lower-frequency twisting/scissoring mode strongly modulates the \textsuperscript{4}T\textsubscript{2g}/\textsuperscript{2}E\textsubscript{g} energy gap, periodically zeroing the energy gap, whereas spin--orbit coupling is essentially invariant to vibrational motion ($\approx$ 60--80\,cm$^{-1}$). Furthermore, calculations of single-point excited-state absorption from \textsuperscript{2}E\textsubscript{g} to a higher ligand-to-metal charge-transfer (LMCT) state indicate that the coherences previously observed in transient absorption spectra arise from nuclear motion on the \textsuperscript{2}E\textsubscript{g} surface. These results provide insights into the correlation between vibrational motion and electronic transitions, which can facilitate rational molecular design of transition metal complexes with desired excited-state properties by leveraging ligand versatility.

\end{abstract}
\section{1. Introduction}

The photochemical reactions of transition metal coordination compounds have attracted much attention because their excited-state dynamics can be systematically tuned by judicious choice of metal centers and ligands. Research on the excited-state dynamics of these complexes aims to find a clear picture of how structure determines electronic transitions and specific relaxation pathways, with the ultimate goal of designing new complexes with desirable properties. The applications of photoexcited transition metal compounds extend to many important application domains, e.g., light-emitting devices, dye-sensitized solar cells, photodynamic therapy, and photochromic materials\cite{costa2012luminescent, polo2004metal, monro2018transition, tian2016photochromic}. 


Tris(acetylacetonato)chromium(III), \ch{Cr(acac)3}, is an archetypical octahedral transition metal complex, which has been broadly studied for its photochemical and photophysical properties.\cite{forster1961luminescence, forster1990photophysics, kirk1999photochemistry, forster2002thermal, forster2006intersystem, kitzmann2022spin} Herein, we select this system as a model system to interrogate the interplay between vibrational motions and intersystem crossing (ISC).   
In contrast to the extensively studied luminescent second- and third-row transition metal complexes with non-innocent ligands (e.g., Ru(II) and Ir(III) polypyridines such as \ch{[Ru(bpy)3]^{2+}} and \ch{Ir(ppy)3})\cite{henry2008early, sun2015role, mishra2025unravelling, yersin2011triplet}, the absence of low-lying metal-to-ligand charge-transfer (MLCT) states facilitates detailed experimental and theoretical analysis. 

\begin{figure}
  \centering
  \includegraphics[width=0.5\linewidth]{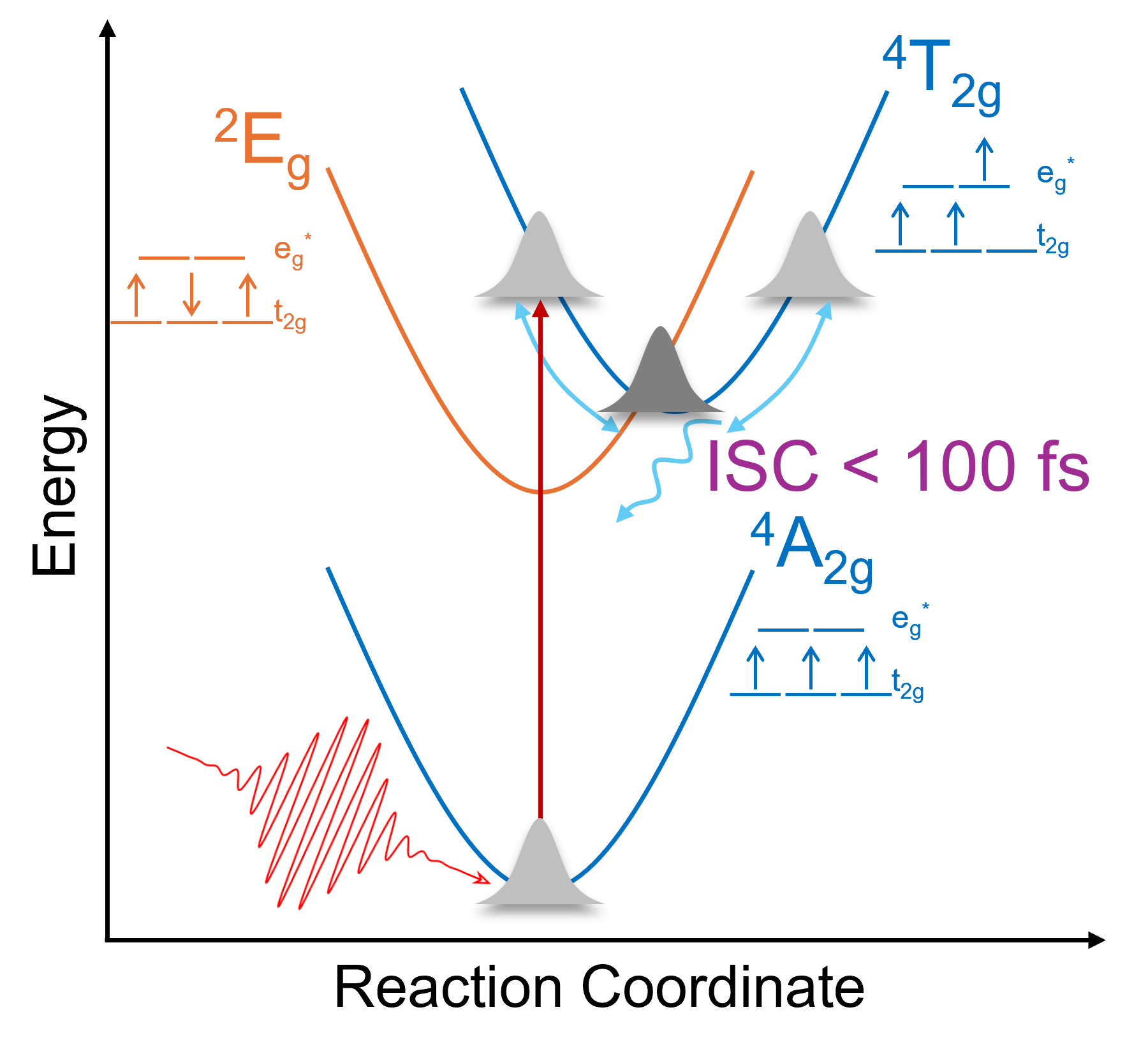}
  \caption{Qualitative PESs of the lowest-lying ligand-field states of \ch{Cr(acac)3} (\textsuperscript{4}A\textsubscript{2g}, \textsuperscript{4}T\textsubscript{2g}, and \textsuperscript{2}E\textsubscript{g}; term symbols in $O_h$ symmetry notation). Photoexcitation (vertical arrow) prepares the wavepacket in the \textsuperscript{4}T\textsubscript{2g} state, which evolves along the reaction coordinate and undergoes intersystem crossing to the \textsuperscript{2}E\textsubscript{g} state in less than 100\,fs near the crossing region. For each electronic state, the one-electron orbital occupations are shown next to the corresponding term symbol to indicate the ligand-field configurations.}
  \label{fig:PES}
\end{figure}

The current mechanistic understanding of \ch{Cr(acac)3} photodynamics has arisen from the combined results of a variety of spectroscopic experiments, e.g., time-resolved infrared spectroscopy, transient absorption (TA) spectroscopy, X-ray absorption spectroscopy, and Raman spectroscopy\cite{maccoas2007relaxation, maccoas2015role, paulus2020insights, paulus2022use, kubin2018cr, ghodrati2025identification}. The lowest three electronic states in the \ch{Cr(acac)3} system are: the \textsuperscript{4}A\textsubscript{2g} ground state quartet, the \textsuperscript{4}T\textsubscript{2g} excited state quartet and the \textsuperscript{2}E\textsubscript{g} excited state doublet.
The ISC of \textsuperscript{4}T\textsubscript{2g} $\rightarrow$ \textsuperscript{2}E\textsubscript{g} transition was reported to occur in less than 100\,fs by the McCusker group in 2005\cite{juban2005ultrafast}, as inferred from the assignments of TA spectra. The ultrafast excited-state dynamics are depicted on the qualitative potential energy surfaces (PESs) of \ch{Cr(acac)3} in Figure~\ref{fig:PES}. Following \textsuperscript{4}A\textsubscript{2g} $\rightarrow$ \textsuperscript{4}T\textsubscript{2g} excitation, vibrational motion of the molecule occurs in the \textsuperscript{4}T\textsubscript{2g} state, and the system subsequently transitions to the lower-lying \textsuperscript{2}E\textsubscript{g} state. A critical feature of this system is that the \textsuperscript{4}T\textsubscript{2g} and \textsuperscript{2}E\textsubscript{g} states are non-parallel, which allows the energy gap to be tuned by specific vibrational modes. Later in 2010, TA spectra by the same research group based on shorter (50\,fs) laser pulses showed vibrational coherences, assigned to dynamics in the \textsuperscript{2}E\textsubscript{g} excited state following ultrafast ISC.\cite{schrauben2010vibrational} This results suggests that the observed vibrational motion at 164\,cm$^{-1}$ can influence the rate of ultrafast ISC. In 2020, further analyses on the TA spectrum of \ch{Cr(acac)3} exhibited two prominent vibrational frequencies, 193\,cm$^{-1}$ and 461\,cm$^{-1}$ respectively\cite{paulus2020insights}. Ultrafast spectroscopy thus provides a means to elucidate the structure--property relationships governing photodynamics and inform the synthetic design of transition metal complexes so that their photochemical properties can be tailored by tuning the ligand-field.

Besides the experimental spectroscopic progress described above, previous theoretical dynamics studies on the ultrafast ISC of \ch{Cr(acac)3} were performed by Ando and co-workers using excited-state wavepacket simulations.\cite{ando2012theoretical} Their work attributed ISC to the \textsuperscript{4}T\textsubscript{2g} $\rightarrow$ \textsuperscript{2}T\textsubscript{1g} transition, driven by crossings of the corresponding PESs and strong spin--orbit couplings (SOCs). The PESs are evaluated along the linear reaction/one-dimensional path primarily consisting of a low-frequency Cr--O vibrational mode that modulates the energy gap between states. 

In the present work, we adopt the Born--Oppenheimer \textit{ab initio} molecular dynamics (BO-AIMD) method with on-the-fly electronic structure calculations in the adiabatic \textsuperscript{4}T\textsubscript{2g} state. This approach explicitly treats all vibrational degrees of freedom throughout the simulation, allowing us to characterize the structural variations and/or specific vibrational motions that facilitate ISC without prior assumptions about which modes are active.  From the theoretical point of view, the exploration of excited-state transition metal complexes poses several challenges. First, their electronic states are intrinsically multiconfigurational: distinct d-orbital configurations and spin multiplicities are nearly degenerate, requiring care in the description of the electronic structure.\cite{ jiang2012multireference, khedkar2021modern}  Multireference approaches are widely considered to be a gold standard \cite{zobel2021quest}. Unfortunately, these complexes are typically large, making multireference treatments costly. As such, a variety of single reference approaches, most often density functional theory, have been widely adopted.\cite{mohr2003use, minenkov2018application,vlvcek2007modeling, buhl2006geometries, niu2000theoretical,niu2000theoretical, hu2000assessment}   
Additionally, it is difficult to definitively assign experimentally observed ultrafast dynamics without directly computing experimental observables.\cite{gelin2021ab, xu2022ultrafast, mehmood2024simulating, silfies2023ultrafast}. Pump-probe TA experiments, like those used to elucidate the dynamics of \ch{Cr(acac)3}, probe the excited-state dynamics through a combination of excited-state absorption (ESA), stimulated emission (SE), and ground-state bleach (GSB).  The ESA features typically originate from transitions from the populated excited state to higher-lying states that are often of charge-transfer character in transition metal complexes. Reliable description of these higher-lying charge-transfer states requires dynamic correlation and incurs higher computational cost.  

In this work, we employ a combination of first-principles molecular dynamics and static electronic structure calculations to explore the relationship between the ultrafast vibrational dynamics of photoexcited \ch{Cr(acac)3} and the ISC process.  To this end, we model the dynamics of a simplified molecular model, tris(1,3-propanedionato)chromium(III), \ch{Cr(PDO)3}, in which the \ch{CH3} groups of \ch{Cr(acac)3} are substituted by hydrogen atoms to reduce computational cost while retaining the same electronic character. We perform Born-Oppenheimer \textit{ab initio} molecular dynamics (BO-AIMD) simulations on the \textsuperscript{4}T\textsubscript{2g} PES and evaluate various electronic properties, including SOCs, along the trajectories to elucidate vibrational motions of nuclei persisting in the excited state. We further identify the doublet ligand-to-metal charge-transfer (\textsuperscript{2}LMCT) state that contributes to the ESA signatures of Cr(III) complexes, allowing us to directly assign the experimentally observed TA signal. In sec. 2, we describe the computational details of our excited-state study. In sec. 3, we present the results and discussion. Finally, we provide the conclusions in Sec. 4.

\section{2. Theoretical Methods}
The ultrafast excited-state dynamics study of Cr(III) complexes mainly includes three components: (i) geometry optimization of the ground and excited states; (ii) AIMD simulations on the \textsuperscript{4}T\textsubscript{2g} surface; and (iii) identification of the doublet LMCT state responsible for the experimentally probed TA signals. The details of each component are described below.

\subsection{2.1 Geometry Optimization for Ground and Excited States}

The geometries of \ch{Cr(acac)3} and \ch{Cr(PDO)3} in selected electronic states are optimized with the Molpro \textit{ab initio} quantum chemistry package.\cite{werner2012molpro, werner2020molpro, werner1985second, knowles1985efficient, kreplin2019second, kreplin2020mcscf, eckert1997ab} We use state-averaged (SA) complete active space self-consistent field (CASSCF) calculations with an active space CAS(3,5) (three d electrons in five d orbitals). To capture the near-degeneracies within each manifold, quartet surfaces are treated with SA-4-CASSCF equally weighted over the \textsuperscript{4}A\textsubscript{2g} ground state and three degenerate \textsuperscript{4}T\textsubscript{2g} excited states, whereas the doublet surface is optimized with SA-2-CASSCF equally weighted over two degenerate \textsuperscript{2}E\textsubscript{g} excited states. For consistency of the d orbitals, CASSCF orbitals optimized for the quartet states are used as the initial guess for the subsequent doublet optimization. We apply the effective core potential basis set LANL2DZ\cite{hay1985ab, wadt1985ab, hay1985ab2}, replacing 10 core electrons of chromium to account for the scalar relativistic effect and increase computational efficiency for transition metal complexes.

\subsection{2.2 \textit{Ab Initio} Molecular Dynamics Method}

To illuminate the role of vibrational dynamics preceding ISC, we apply the BO-AIMD method, and we can better comprehend ISC by tracing the SOCs and energy gaps along the trajectories. The excited-state dynamics of \ch{Cr(PDO)3} are simulated using this method as implemented in PySpawn,\cite{fedorov2020pyspawn} which is interfaced to the TeraChem quantum chemistry software package for graphics processing unit-accelerated electronic structure calculations.\cite{seritan2020terachem, seritan2021terachem, ufimtsev2009quantum, titov2013generating, song2016automated, fales2017complete, hohenstein2015atomic, snyder2015atomic}. Pyspawn is a Python implementation of the \textit{ab initio} multiple spawning molecular dynamics method,\cite{ben2000ab} but in this work, spawning of trajectory basis functions is disabled so that the trajectories evolve adiabatically on the first excited quartet state \textsuperscript{4}T\textsubscript{2g}. All trajectories of dynamics simulations employ the SA-4-CAS(3,5)SCF electronic structure method. A time step of 15\,a.u. ($\sim$0.36\,fs) is adopted for a total simulation length of 25200\,a.u. ($\sim$610\,fs). The 50 independent trajectories are propagated to obtain a representative ensemble for the \textsuperscript{4}T\textsubscript{2g} excited-state dynamics. The initial conditions, including positions and momenta, are sampled from the Wigner distribution\cite{wigner1932quantum} generated at the ground state equilibrium geometry within the harmonic approximation at SA-4-CAS(3,5)SCF\slash LANL2DZ level of theory. Sampling from the Wigner distribution provides classical trajectories with quantum initial positions of nuclei. All the starting geometries are prepared from the ground state and instantaneously promoted to the \textsuperscript{4}T\textsubscript{2g} state within the Franck--Condon region, mimicking excitation by a pump pulse after which the excited-state dynamics proceed. The BO-AIMD simulations are performed in the gas phase without solvent molecules.

\subsection{2.3 Identification of \textsuperscript{2}LMCT States}

Our BO-AIMD simulations are designed to elucidate the initial dynamics on the \textsuperscript{4}T\textsubscript{2g} surface. However, the experimental TA signal that exhibits vibronic coherence has been assigned to the ESA from the \textsuperscript{2}E\textsubscript{g} state to a higher-lying doublet state. This is because the ultrafast ISC from \textsuperscript{4}T\textsubscript{2g} to \textsuperscript{2}E\textsubscript{g} appears to occur within 100\,fs, so the experimentally observed subpicosecond vibronic coherence ultimately resides on the \textsuperscript{2}E\textsubscript{g} surface. The experimentally-observed ESA therefore is expected to reflect the energy gap changes between \textsuperscript{2}E\textsubscript{g} and a higher-lying doublet state as the nuclear wavepacket oscillates on the \textsuperscript{2}E\textsubscript{g} surface. This higher-lying doublet state is expected to have a \textsuperscript{2}LMCT character because the acac$^-$ ring is a $\pi$-donor with relatively high-lying doubly occupied 2p orbitals of oxygen, while the low-lying $t_{2g}$ orbitals of Cr(III) are acceptor orbitals singly occupied. Consequently, the lowest-energy charge-transfer pathway is LMCT rather than MLCT. This spin- and symmetry-allowed \textsuperscript{2}E\textsubscript{g} $\rightarrow$ \textsuperscript{2}LMCT transition is expected to exhibit a large transition dipole moment.

To characterize the excitation of the doublet, we employ extended dynamically weighted (XDW) CASPT2 as implemented in the OpenMolcas package\cite{fdez2019openmolcas, battaglia2020extended}, evaluating vertical excitation energies at two representative geometries: the ground state equilibrium geometry and the \textsuperscript{4}T\textsubscript{2g} minimum. To capture the charge-transfer character, the active space is expanded to include three additional ligand $\pi$ orbitals, yielding a CAS(9,8) (9 electrons in 8 orbitals) for the doublet excited states. Orbitals are optimized by state-averaging over 41 doublet states with equal weights (SA-41), ensuring a balanced description of higher-lying LMCT configurations. We initialize orbital optimization with SA-41-CAS(3,5)SCF and expand to SA-41-CAS(9,8)SCF, which provides the CASSCF reference for the subsequent XDW-CASPT2 calculation. We use the def2-SVP basis set\cite{weigend2005balanced} for these correlated wavefunction calculations and add scalar relativistic effects via the X2C (exact-two-component) corrections\cite{liu2009exact}. The ionization potential electron affinity shift is set to 0 hartree, and an imaginary shift of 0.25 hartree is applied to mitigate intruder states\cite{ghigo2004modified, forsberg1997multiconfiguration}. The 20 core electrons of \ch{Cr(PDO)3} are frozen to reduce cost. The \textsuperscript{2}LMCT state is identified as the first doublet with dominant LMCT character, i.e., a leading configuration corresponding to an electron promotion from a ligand $\pi$ orbital to a Cr(III) $t_{2g}$ orbital.

\section{3. Results and Discussion}
\subsection{3.1 Validating the Ligand-Field States: Metal-Ligand Bond Lengths}

\begin{figure}
  \centering
  \includegraphics[width=0.6\textwidth]{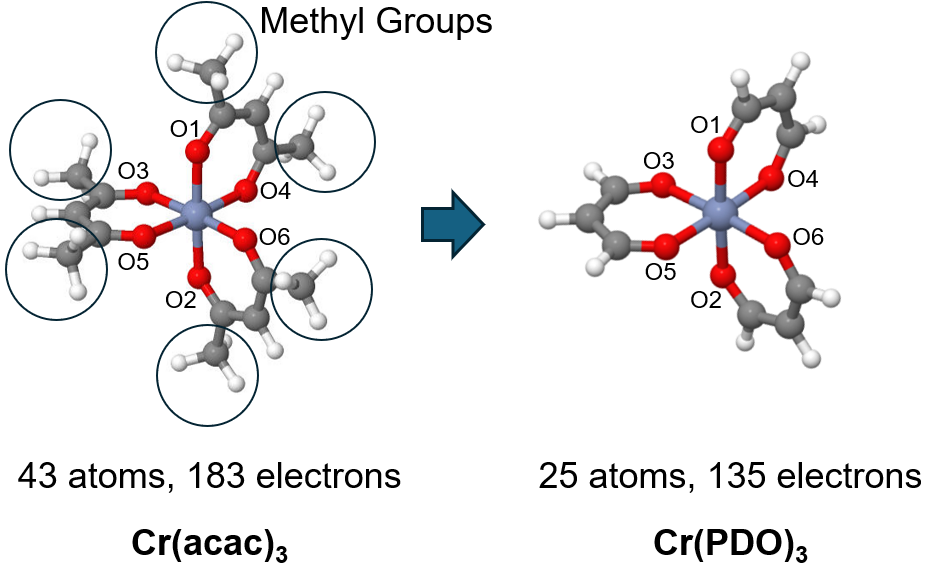}
  \caption{Structures of \ch{Cr(acac)3} and the truncated model \ch{Cr(PDO)3} used in this work (methyl $\rightarrow$ H substitution) to lower the computational cost by reducing the number of atoms and electrons while preserving the inner coordination sphere and ligand-field excited states.}
  \label{fig:Cr(III) system model}
\end{figure}

As shown in Figure~\ref{fig:Cr(III) system model}, we replace methyl substituents with hydrogen atoms, reducing the system from 43 to 25 atoms and from 183 to 135 electrons, thereby lowering the computational cost of the subsequent excited-state dynamics simulations. The inner coordination sphere and the ligand-field splitting of the d orbitals are expected to be preserved in \ch{Cr(PDO)3}, retaining the same excited states. In both complexes, the \textsuperscript{4}A\textsubscript{2g} minimum features six nearly equivalent Cr--O bond lengths of 1.98~\AA, consistent with octahedral symmetry and in good agreement with the crystallographic structure (within a few hundredths of an angstrom; Tables~\ref{tab:Cr(PDO)3 bond lengths} and~\ref{tab:Cr(acac)3 bond lengths}). The \textsuperscript{2}E\textsubscript{g} minimum is geometrically similar to \textsuperscript{4}A\textsubscript{2g} one, consistent with an intraconfigurational spin flip. By contrast, the \textsuperscript{4}T\textsubscript{2g} minimum in both complexes exhibits a pronounced Jahn--Teller (JT) distortion/effect\cite{jahn1937stability, bersuker2001modern, halcrow2013jahn} that reduces symmetry and lifts the electronic degeneracy, characterized by two axial bonds shortened and four equatorial ones elongated relative to the \textsuperscript{4}A\textsubscript{2g} minimum.

\renewcommand{\arraystretch}{1.3} 

\begin{table}[h!]
  \centering
  \begin{tabular}{ccccc}
  \hline
     & \textbf{\textsuperscript{4}A\textsubscript{2g}} & \textbf{\textsuperscript{4}T\textsubscript{2g}} & \textbf{\textsuperscript{2}E\textsubscript{g}} & \textbf{Exp.(GS)}\cite{glick1975tris} \\ \hline
    Cr--O(1) & 1.983 & 1.960 & 1.979 & 1.962 \\
    Cr--O(2) & 1.984 & 1.957 & 1.979 & 1.945 \\
    Cr--O(3) & 1.984 & 2.010 & 1.984 & 1.954 \\
    Cr--O(4) & 1.984 & 2.064 & 1.985 & 1.961 \\
    Cr--O(5) & 1.984 & 2.170 & 1.983 & 1.955 \\
    Cr--O(6) & 1.984 & 2.035 & 1.985 & 1.931 \\ \hline
  \end{tabular}
  \caption{Optimized geometries of \ch{Cr(PDO)3} in the \textsuperscript{4}A\textsubscript{2g}, \textsuperscript{4}T\textsubscript{2g}, and \textsuperscript{2}E\textsubscript{g} states, shown alongside the crystallographic ground state (GS) equilibrium geometry for comparison. Structures are characterized by Cr--O bond lengths.}
  \label{tab:Cr(PDO)3 bond lengths}
\end{table}

\begin{table}[h!]
  \centering
  \begin{tabular}{ccccc}
  \hline
     & \textbf{\textsuperscript{4}A\textsubscript{2g}} & \textbf{\textsuperscript{4}T\textsubscript{2g}} & \textbf{\textsuperscript{2}E\textsubscript{g}} & \textbf{Exp.(GS)}\cite{morosin1965crystal} \\ \hline
    Cr--O(1) & 1.980 & 1.954 & 1.975 & 1.943 \\
    Cr--O(2) & 1.979 & 1.955 & 1.975 & 1.942 \\
    Cr--O(3) & 1.980 & 2.097 & 1.978 & 1.951 \\
    Cr--O(4) & 1.979 & 2.037 & 1.981 & 1.958 \\
    Cr--O(5) & 1.979 & 2.065 & 1.980 & 1.956 \\
    Cr--O(6) & 1.979 & 2.044 & 1.980 & 1.959 \\ \hline
  \end{tabular}
  \caption{Optimized geometries of \ch{Cr(acac)3} in the \textsuperscript{4}A\textsubscript{2g}, \textsuperscript{4}T\textsubscript{2g}, and \textsuperscript{2}E\textsubscript{g} states, shown alongside the crystallographic ground state (GS) equilibrium geometry for comparison. Structures are characterized by Cr--O bond lengths.}
  \label{tab:Cr(acac)3 bond lengths}
\end{table}

\subsection{3.2 Ultrafast Excited-State Dynamics on the \textsuperscript{4}T\textsubscript{2g} Surface}

In this work, we focus on the \textsuperscript{4}T\textsubscript{2g} excited-state dynamics that initiate the ultrafast ISC.
To this end, we employ the BO-AIMD method to explicitly resolve the Cr--O bond stretching dynamics on the \textsuperscript{4}T\textsubscript{2g} surface and the associated time-dependence of key electronic structure properties.

\subsection{3.2.1 Metal-Ligand Bond-Length Dynamics and Vibrational Motions}

\begin{figure}
  \centering
  \includegraphics[width=0.6\linewidth]{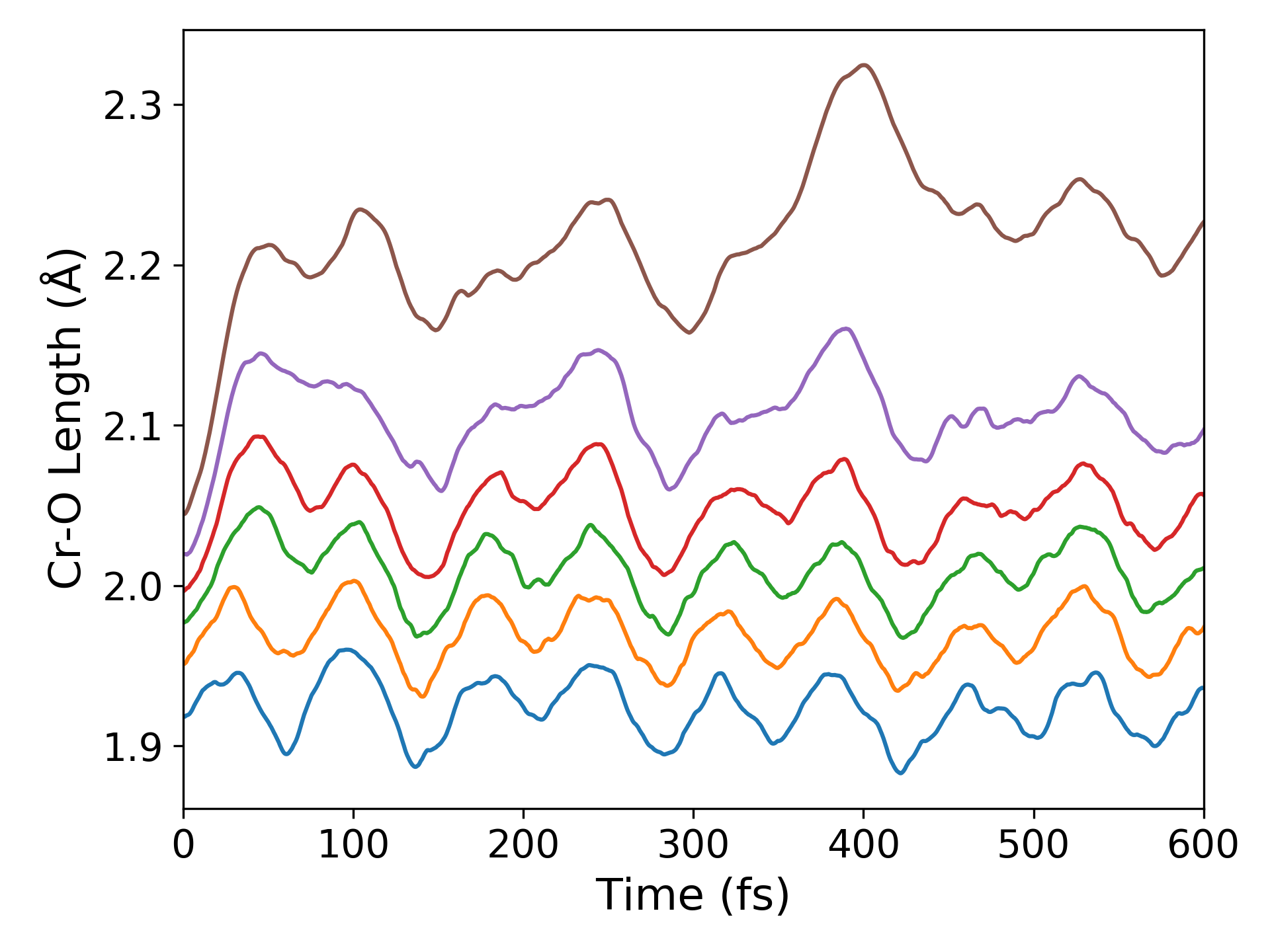}
  \caption{The sorted Cr--O bond-length dynamics on \textsuperscript{4}T\textsubscript{2g} surface. The six Cr--O bond lengths are ranked by length at each time step so curves show the order statistics (shortest $\rightarrow$ longest) averaged over 50 trajectories instead of tracking individually the specific bond. The average Cr--O bond length at the \textsuperscript{4}A\textsubscript{2g} minimum is 1.98~\AA. Oscillations indicate an active stretching of metal-ligand bonds and a persistent JT distortion on the \textsuperscript{4}T\textsubscript{2g} surface.}
  \label{fig:Average bond length}
\end{figure}

\begin{figure}
  \centering
  \includegraphics[width=0.6\linewidth]{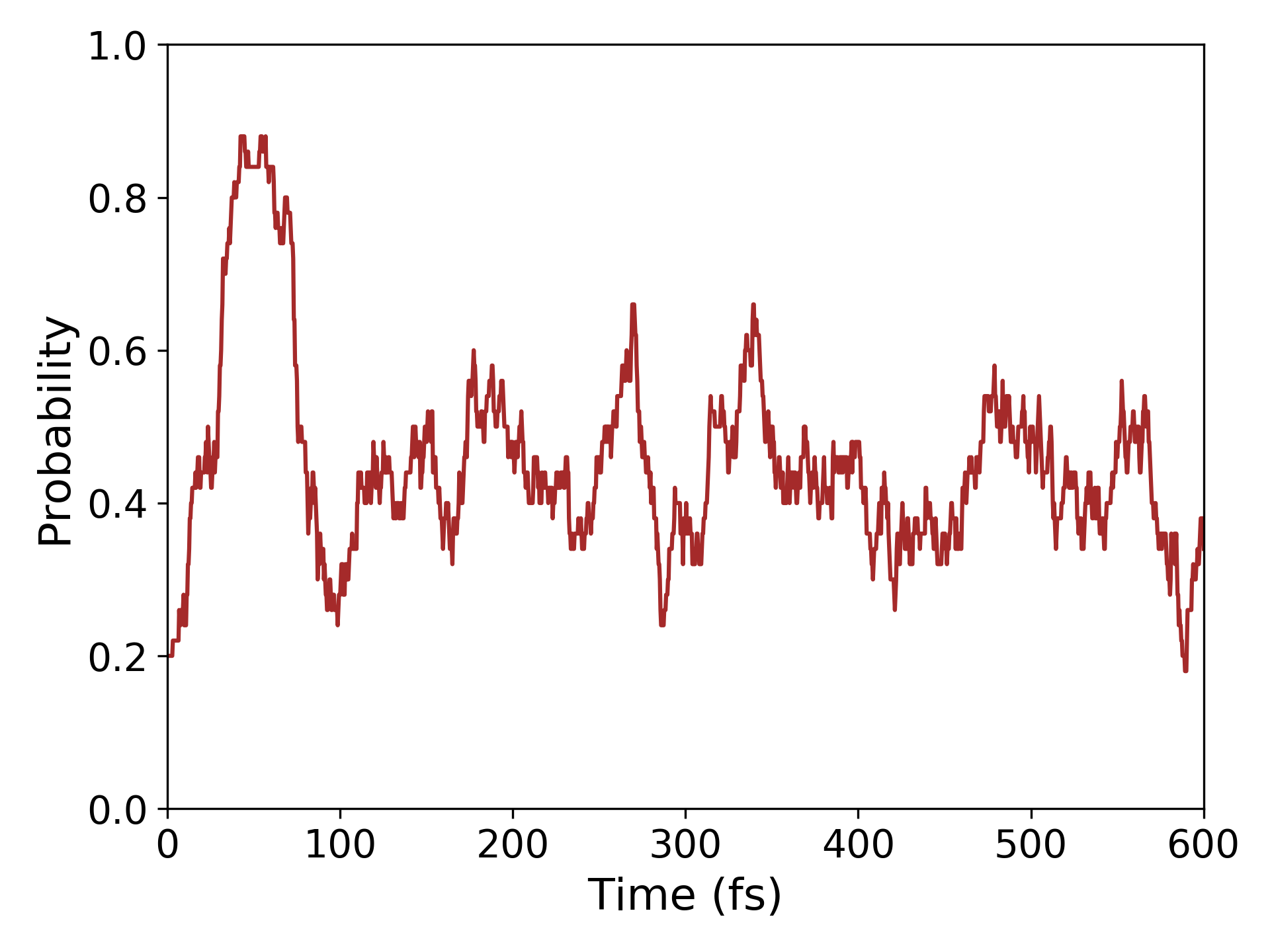}
  \caption{The JT distortion excited-state dynamics. Each geometry is given a one-hot label after the dynamics simulation (1 if the two shortest Cr--O bonds are opposite, 0 otherwise), so the probability is defined as the mean of this label among trajectories.}
  \label{fig:JT bond dynamics}
\end{figure}

The BO-AIMD simulations yield the time evolution of all nuclear coordinates, allowing us to track structural dynamics in the octahedral complex. We analyze the Cr--O bond-length dynamics because this coordinate is potentially coupled to the electronically excited states. As shown in Figure~\ref{fig:Average bond length}, the six Cr--O bonds undergo coherent oscillations between 1.9~\AA\ and 2.3~\AA, indicative of an active stretching mode. At each time step, the bonds are ordered by length and, because they are symmetry-equivalent, the curves plot the instantaneous distribution (the shortest to the longest) averaged over 50 trajectories rather than individual bond identities. For most parts of the simulation, four of the six bonds are longer than 1.98~\AA\ (the average Cr--O distance at the \textsuperscript{4}A\textsubscript{2g} minimum) while two are shorter, consistent with a persistent JT distortion on the \textsuperscript{4}T\textsubscript{2g} surface. Figure~\ref{fig:JT bond dynamics} further quantifies the JT distortion by reporting the probability that the two shortest Cr--O bonds are opposite to each other as a function of time. For an undistorted octahedron, this probability is 0.2 (i.e., 1/5). For each geometry at each time step in each trajectory, we assign a one-hot label (1 if the two shortest bonds are opposite, 0 otherwise); the time-dependent probability is then the ensemble mean of this label across all the trajectories at each time step. Initiated from the Franck--Condon region near the ground state equilibrium geometry, the probability starts near 0.2 at t = 0, rises rapidly to > 0.8 by $\sim$50\,fs, and subsequently oscillates around $\sim$0.4---remaining above 0.2 and thereby demonstrating sustained JT character on the \textsuperscript{4}T\textsubscript{2g} surface.

\begin{figure}
  \centering
  \includegraphics[width=0.6\linewidth]{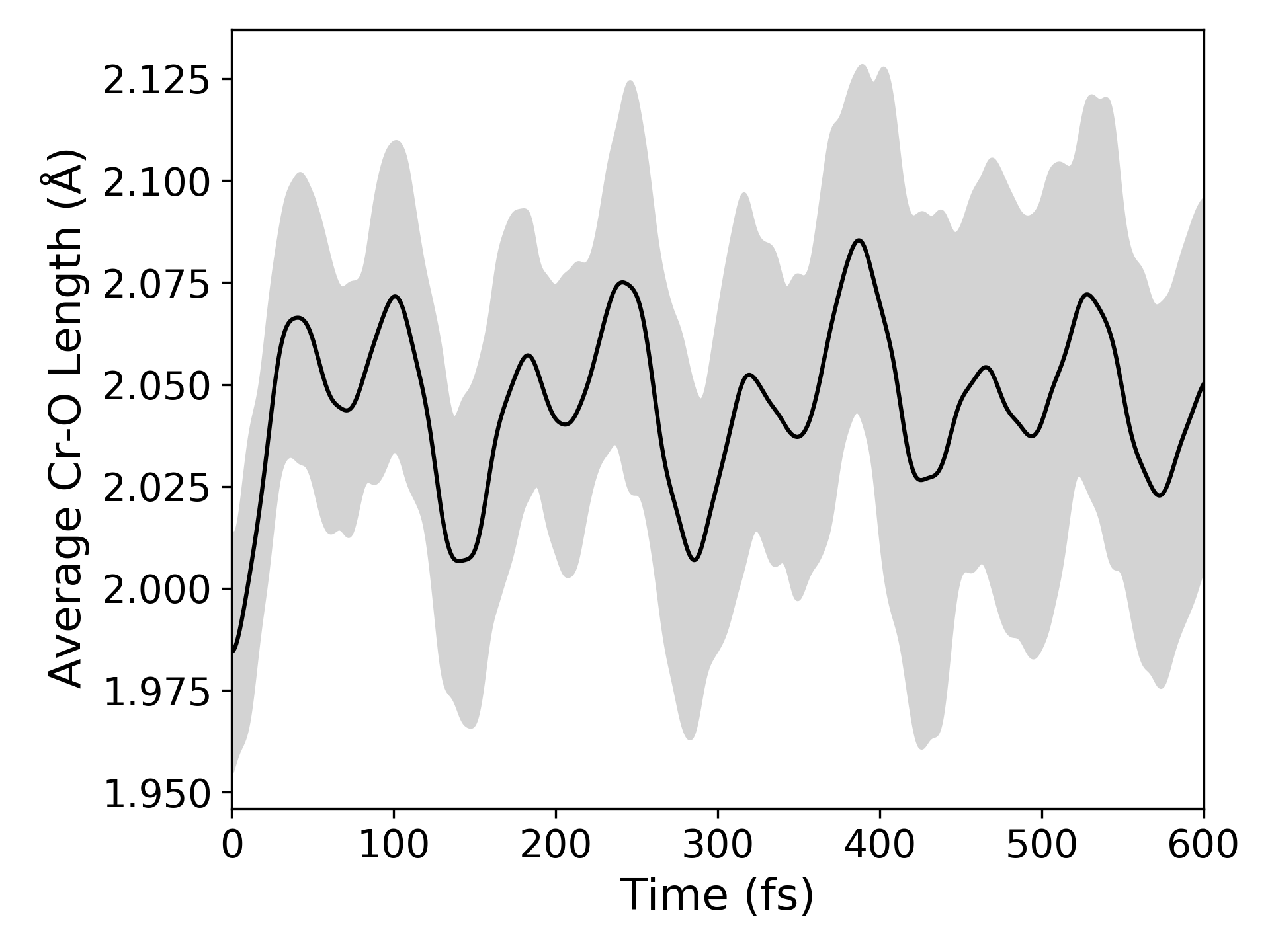}
  \caption{The bond dynamics of symmetric Cr--O stretching. The trace shows the mean Cr--O bond length, averaged over all Cr--O bonds within each geometry and across 50 trajectories. The gray band represents $\pm\,2\sigma$. The time-averaged standard deviation is 0.02~\AA. The oscillatory pattern exhibits a coherent symmetric nuclear motion.}
  \label{fig:Sym bond dynamics}
\end{figure}

\begin{figure}
  \centering
  \includegraphics[width=0.6\linewidth]{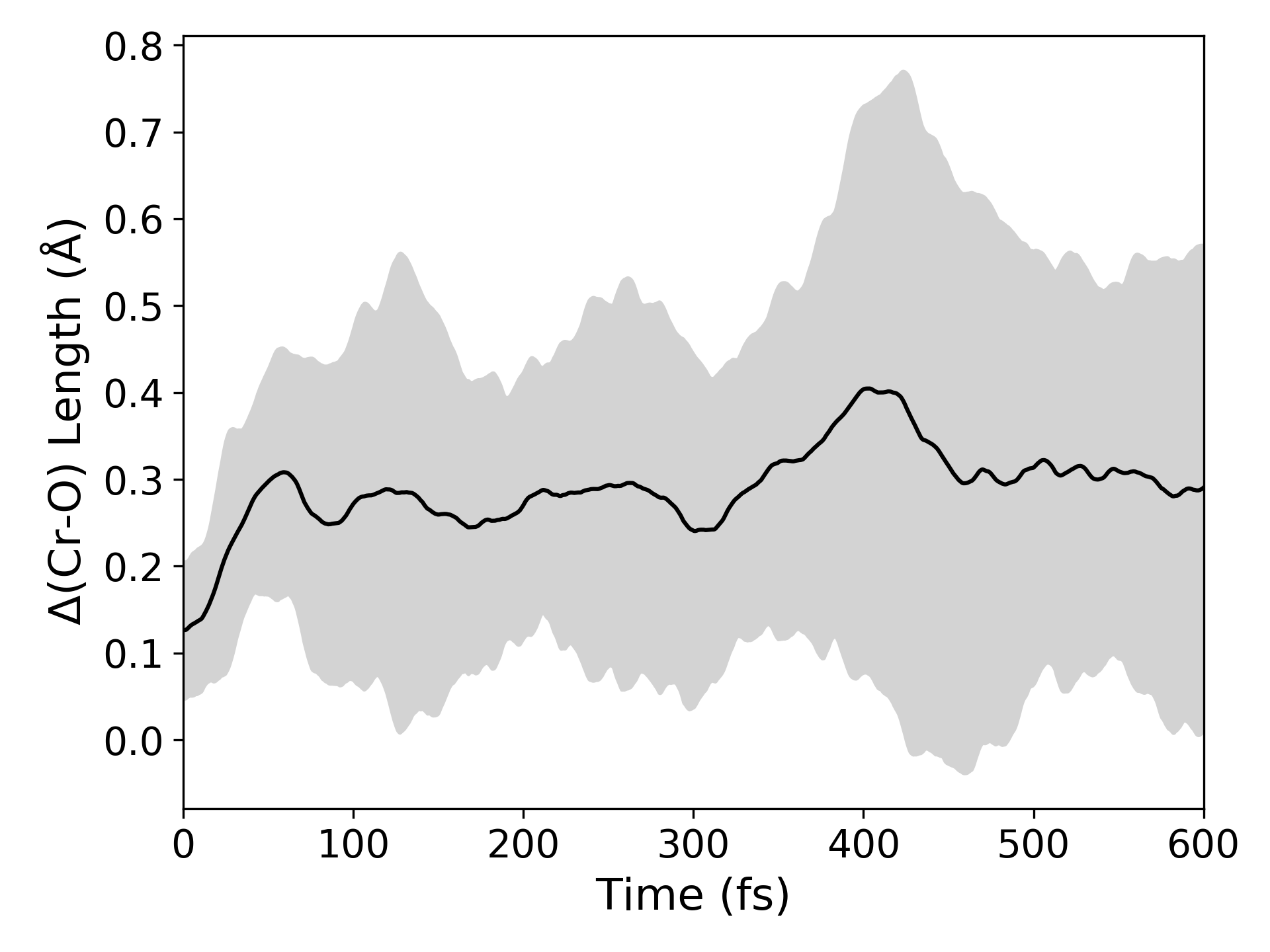}
  \caption{The bond dynamics of asymmetric Cr--O stretching. The trace reports the instantaneous difference between the longest and shortest Cr--O bonds, ensemble-averaged over 50 trajectories. The gray band denotes $\pm\,2\sigma$. The time-averaged standard deviation is 0.11~\AA, indicating a largely incoherent asymmetric nuclear motion.}
  \label{fig:Asy bond dynamics}
\end{figure}

To further interrogate the Cr--O stretching dynamics, we decompose the motion into symmetric and asymmetric components. The symmetric motion corresponds to in-phase stretching of all six Cr--O bonds, whereas the asymmetric motion captures antiphase distortions among the six bonds. In Figure~\ref{fig:Sym bond dynamics}, the black curve shows the mean Cr--O bond length, obtained by averaging over the six bonds within each configuration and then over 50 trajectories (i.e., the totally symmetric stretching motion). It exhibits coherent femtosecond-scale oscillations between 2.00~\AA\ and 2.10~\AA. The gray band denotes $\pm\,2$ standard deviations ($\pm\,2\sigma$), indicating a modest inter-trajectory spread; the time-averaged standard deviation is 0.02~\AA. On the contrary, the asymmetric stretching motion, quantified as the trajectory-averaged difference between the longest and shortest Cr--O bonds at each time step (Figure~\ref{fig:Asy bond dynamics}), displays incoherent oscillations. Because asymmetric distortions vary among trajectories, they largely cancel out in the ensemble average while the dispersion remains substantial (gray band denotes $\pm\,2\sigma$). The standard deviation of the bond difference fluctuates strongly over the simulation and has a time-averaged value of 0.11~\AA, fivefold larger than that of the symmetric stretching.

\begin{figure}
  \centering
  \includegraphics[width=0.6\linewidth]{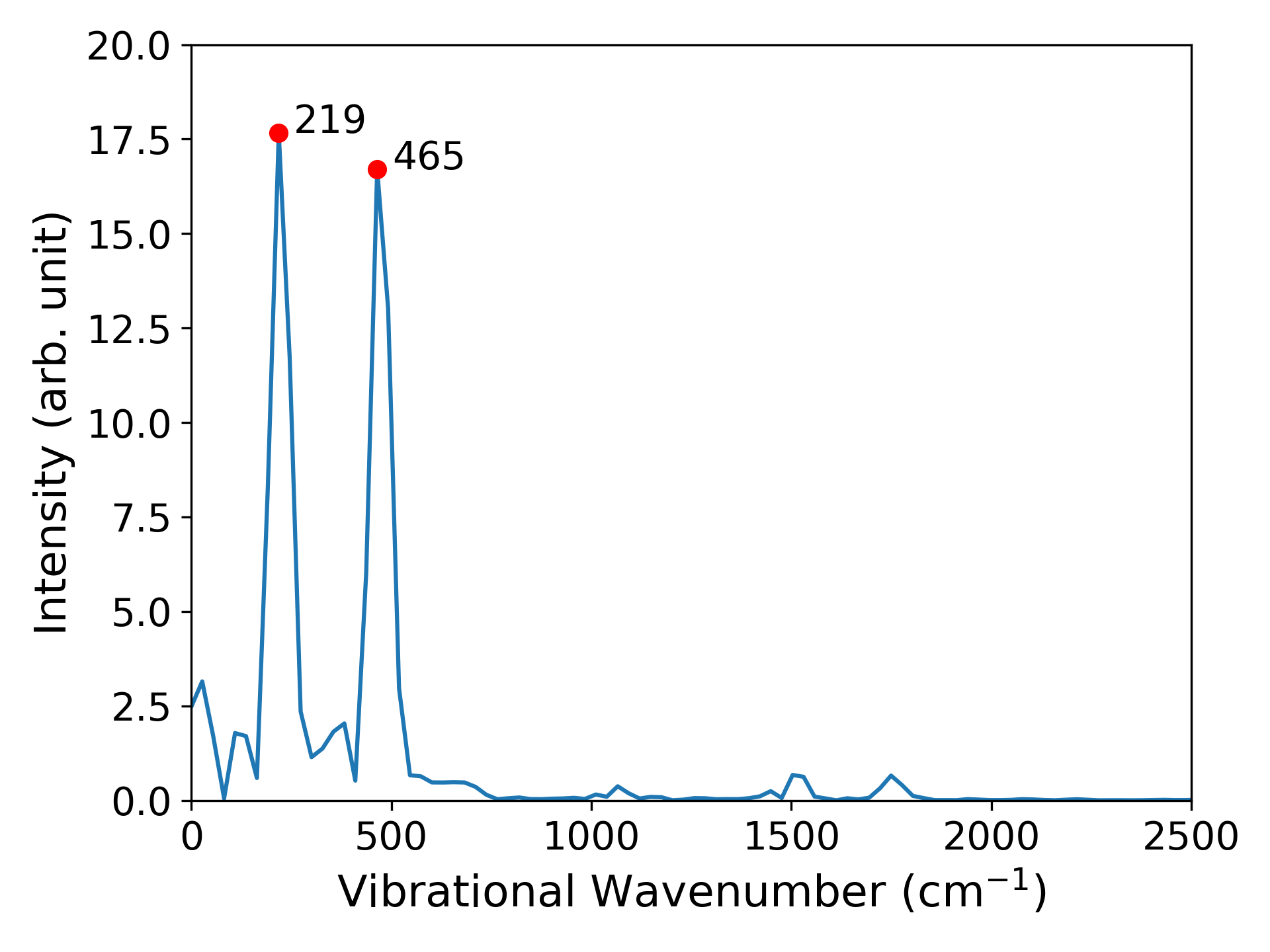}
  \caption{Fourier transform analysis of the mean Cr--O bond length trace; see text for data preprocessing details. The spectral resolution is 27\,cm$^{-1}$.}
  \label{fig:FFT bonds}
\end{figure}

Given the coherence of the symmetric stretching motion and its clear oscillatory pattern in the average Cr--O bond length, we compute the Fourier transform (FT) of the time trace to obtain the vibrational frequencies (Figure~\ref{fig:FFT bonds}). Prior to FT, the signal from Figure~\ref{fig:Sym bond dynamics} is mirrored about t = 0, mean-subtracted, and tapered with a Hanning window to mitigate endpoint discontinuities and spectral leakage. The spectral resolution is 27\,cm$^{-1}$, set by the total simulation time (inversely proportional to total simulation time). The FT reveals two dominant peaks at 219\,cm$^{-1}$ and 465\,cm$^{-1}$, corresponding to oscillation periods of 152\,fs and 72\,fs, respectively. These features are in good agreement with experimental frequencies of 193\,cm$^{-1}$ and 461\,cm$^{-1}$ in the Ref.~\citenum{paulus2022use}, converted from coherent oscillations on the TA spectrum. On the basis of different dephasing times, the 193\,cm$^{-1}$ mode was further assigned as the reaction coordinate for ISC, whereas the 461\,cm$^{-1}$ mode was identified as a purely vibrational mode.

\begin{figure}
  \centering
  \includegraphics[width=0.8\linewidth]{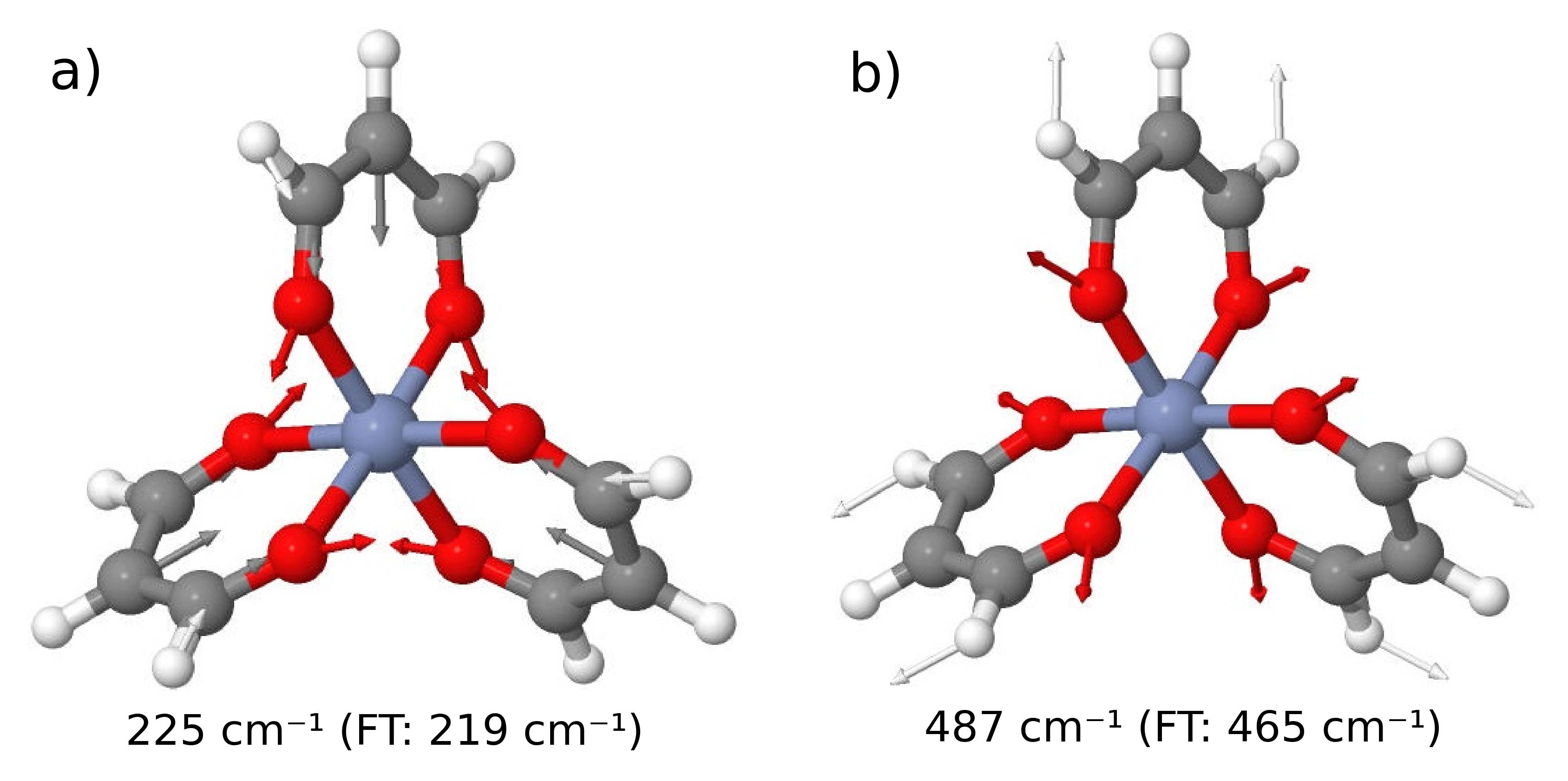}
  \caption{Normal modes associated with the symmetric Cr--O stretching motion: a) 225\,cm$^{-1}$ (twisting/scissoring with global symmetric breathing mode) and b) 487\,cm$^{-1}$ (predominantly collinear Cr--O stretching mode). Ground state vibrational frequencies are computed at the SA-4-CAS(3,5)SCF\slash LANL2DZ level. The corresponding Fourier transform (FT) peak frequencies, derived from the excited-state dynamics data, are given in brackets.}
    \label{fig:two modes}
\end{figure}

We next perform a ground state vibrational frequency analysis at SA-4-CAS(3,5)SCF\slash LANL2DZ level of theory. Two assumptions are made: (i) a harmonic approximation for the \textsuperscript{4}A\textsubscript{2g} and \textsuperscript{4}T\textsubscript{2g} states, and (ii) minimal changes in normal modes between these states. We manually identify two normal modes with a dominant symmetric Cr--O stretching character. The lower-frequency mode at 225\,cm$^{-1}$ (Figure~\ref{fig:two modes}a) involves primarily tangential motion (with respect to the Cr--O axes) of the six O atoms; three pairs of Cr--O bonds undergo a synchronous twisting/scissoring motion that produces a breathing-like expansion and contraction of the \ch{acac-} rings. The higher-frequency mode at 487\,cm$^{-1}$ (Figure~\ref{fig:two modes}b) features motion largely colinear with the Cr--O axes, with in-phase bond lengthening/shortening and only small displacements of the chelate backbones (except for the immobile central C atoms). Both modes are in reasonable agreement with the frequencies obtained from the FT of the symmetric Cr--O stretching dynamics. The role of these two modes in the ISC process will be further examined below.

\subsection{3.2.2 Excited-State Energy Dynamics and Role of SOC}

\begin{figure}
  \centering
  \includegraphics[width=0.6\linewidth]{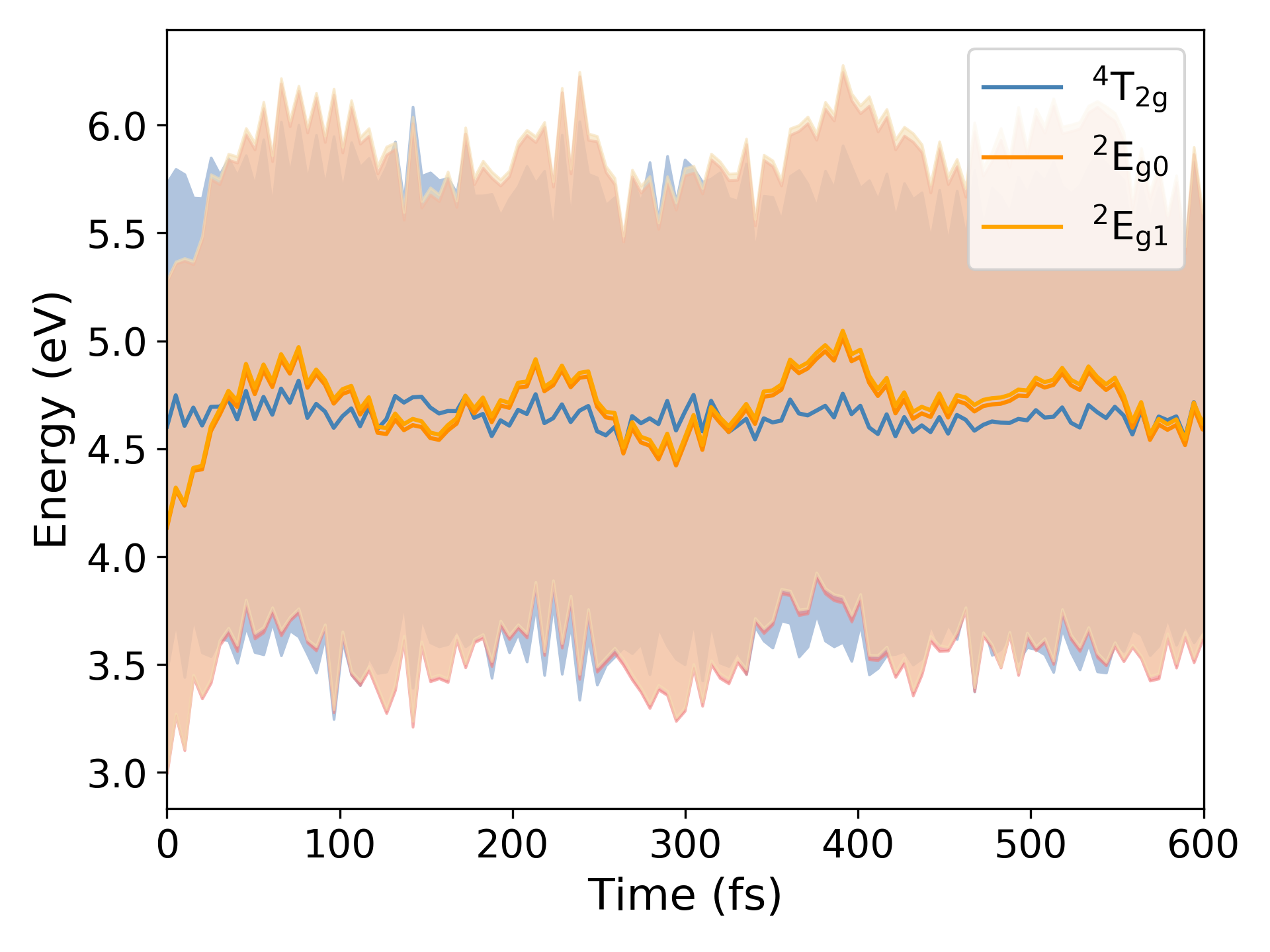}
  \caption{Dynamics of excited-state energies. The energies of the lowest \textsuperscript{4}T\textsubscript{2g} state (blue line) and the two nearly degenerate states of \textsuperscript{2}E\textsubscript{g}, \textsuperscript{2}E\textsubscript{g0} and \textsuperscript{2}E\textsubscript{g1} (orange lines), are shown. Shaded regions indicate the corresponding $\pm\,2\sigma$. The ground state minimum is set to 0. Energies of \textsuperscript{4}T\textsubscript{2g} and \textsuperscript{2}E\textsubscript{g} are shifted to align their respective absorption and emission with experiments.}
  \label{fig:Energies dynamics}
\end{figure}

Figure~\ref{fig:Energies dynamics} shows the time evolution of the energies of \textsuperscript{4}T\textsubscript{2g} and \textsuperscript{2}E\textsubscript{g} ligand-field states along the BO-AIMD trajectories, with the ground state minimum set to zero. To place the excited-state energies on an experimental scale, constant shifts are applied so that the calculated \textsuperscript{4}A\textsubscript{2g} $\rightarrow$ \textsuperscript{4}T\textsubscript{2g} absorption (1.84\,eV) at SA-4-CAS(3,5)SCF\slash LANL2DZ level and \textsuperscript{2}E\textsubscript{g} $\rightarrow$ \textsuperscript{4}A\textsubscript{2g} emission (2.36\,eV) at SA-2-CAS(3,5)SCF\slash LANL2DZ level match with the experimental transition energies of 17700\,cm$^{-1}$ (2.19\,eV)\cite{atanasov1987role} and 12942\,cm$^{-1}$ (1.60\,eV)\cite{schonherr1983hochaufgeloste}, respectively. The shaded bands denote $\pm\,2\sigma$, indicating inter-trajectory variability. The energy of \textsuperscript{4}T\textsubscript{2g} is nearly stationary throughout the simulation, whereas two nearly degenerate states of \textsuperscript{2}E\textsubscript{g} (\textsuperscript{2}E\textsubscript{g0} and \textsuperscript{2}E\textsubscript{g1}) oscillate with a period of $\sim$154\,fs, commensurate with the lower-frequency symmetric Cr--O stretching. Hence, the \textsuperscript{2}E\textsubscript{g} state is more susceptible to nuclear motion than the \textsuperscript{4}T\textsubscript{2g} state, and in particular the 225\,cm$^{-1}$ twisting/scissoring mode couples strongly to the \textsuperscript{2}E\textsubscript{g} electronic state. Notably, the \textsuperscript{4}T\textsubscript{2g} and \textsuperscript{2}E\textsubscript{g} states attain degeneracy at several time steps, indicating access to surface crossings between the two PESs. Such surface crossings provide efficient pathways for the nonradiative relaxation. Accordingly, the 225\,cm$^{-1}$ mode likely promotes ISC by modulating the \textsuperscript{2}E\textsubscript{g} energy and steering the system towards these surface crossings. This picture is also consistent with prior discussion in Ref.~\citenum{paulus2022use}, which implicates the significance of twisting motion in the ISC of several related Cr(III) and Fe(II) transition metal complexes.   It is also consistent with the assignment of the vibrational coherences experimentally observed in Ref.~\citenum{schrauben2010vibrational} to coherent vibrational motion that originates in the \textsuperscript{4}T\textsubscript{2g} state, is sustained through ISC, and continues in the \textsuperscript{2}E\textsubscript{g} state. Evidence for strong coupling of this low-frequency mode to ultrafast ISC was also reported for a \textit{tert}-butyl substituted \ch{Cr(acac)3} derivative\cite{schrauben2010vibrational}, although a later reassessment cautioned that scatter in dephasing times precludes a definitive conclusion.  Generally, increasing ligand sterics dampens ligand motions and can thereby lower the ISC rate.

\begin{figure}
  \centering
  \includegraphics[width=0.6\linewidth]{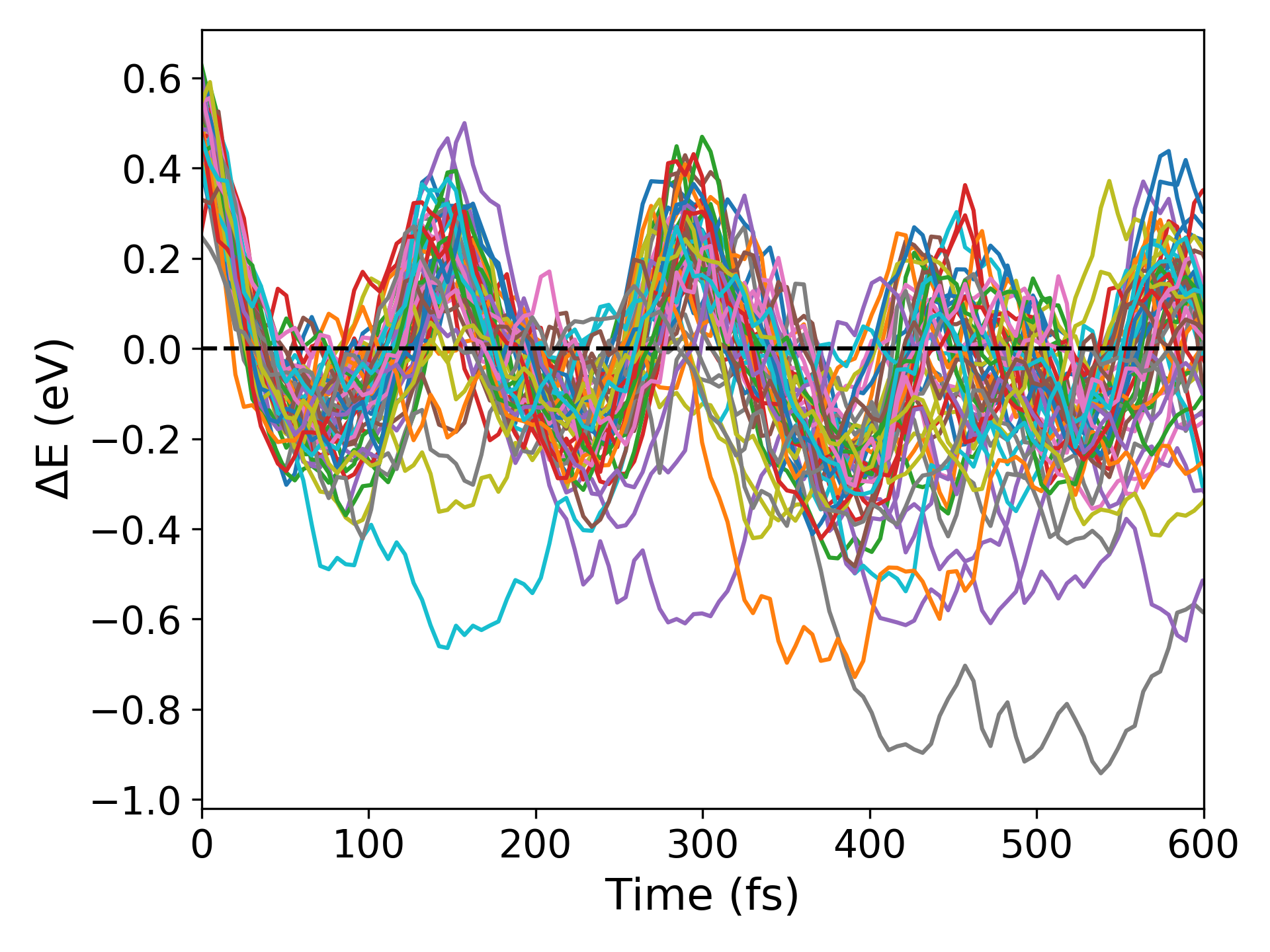}
  \caption{Dynamics of the \textsuperscript{4}T\textsubscript{2g}--\textsuperscript{2}E\textsubscript{g} energy gap, $\Delta$E = E(\textsuperscript{4}T\textsubscript{2g}) -- E(\textsuperscript{2}E\textsubscript{g0}). Each colored curve represents one of 50 trajectories. The horizontal dashed line marks $\Delta$E = 0\,eV (degeneracy).}
  \label{fig:Energy difference dynamics}
\end{figure}

To exclude the possibility that apparent crossings arise as an ensemble-average artifact, we analyze trajectory-resolved time traces of the \textsuperscript{4}T\textsubscript{2g}--\textsuperscript{2}E\textsubscript{g} energy gap (taking the lowest \textsuperscript{2}E\textsubscript{g0} state as a representative doublet; Figure~\ref{fig:Energy difference dynamics}). For most trajectories, the gap exhibits coherent oscillations and repeatedly crosses zero, confirming access to surface crossings along the vibrational coordinates. These results partially address the question posed by McCusker and co-workers (final sentence of the subsection "The basic photophysics of \ch{Cr(acac)3}" in Ref.~\citenum{paulus2022use}) concerning the extent to which the dephasing times of vibrational modes indicate ISC involvement: while dephasing times are suggestive, direct AIMD simulations provide a complementary, mechanistic means to identify the modes that couple to ISC and thus guide strategies to tune the ultrafast excited-state dynamics.

\begin{figure}
  \centering
  \includegraphics[width=0.6\linewidth]{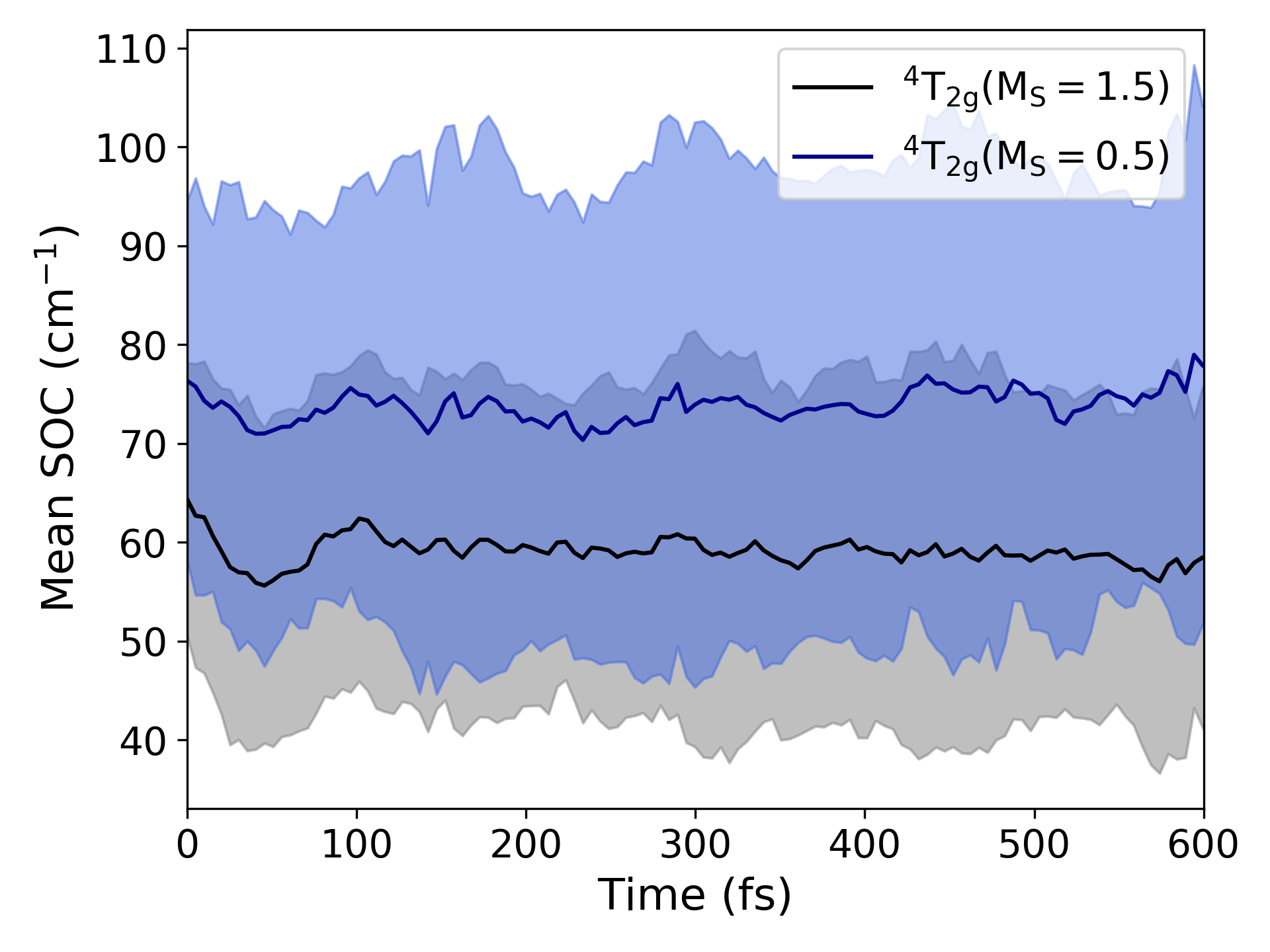}
  \caption{Dynamics of SOCs between the lowest energy \textsuperscript{4}T\textsubscript{2g} quartet and the \textsuperscript{2}E\textsubscript{g} doublets. The SOC matrix elements are grouped by the quartet spin projection M\textsubscript{S} (black: M\textsubscript{S} = $1.5$; blue: M\textsubscript{S} = $0.5$) and reported as the mean SOC magnitude (eq~\ref{eq:soc_mean}). Calculations employ the Breit-Pauli operator. By SOC matrix symmetry, results for $\pm$ M\textsubscript{S} are identical (e.g., values of M\textsubscript{S} = $\pm\,1.5$ are equal). Shaded areas indicate the corresponding $\pm\,2\sigma$.}
  \label{fig:Soc dynamics}
\end{figure}

Another factor that can influence the ISC process is SOC between the \textsuperscript{4}T\textsubscript{2g} and \textsuperscript{2}E\textsubscript{g} states. Figure~\ref{fig:Soc dynamics} reports the time evolution of the mean SOC magnitude, defined as
\begin{equation}
\overline{|H_{\text{SOC}}|}(\mathrm{M_S^Q}) \;=\; \frac{1}{4}\sum_{i=1}^{4} \left| \langle\, \mathrm{Q},\, \mathrm{M_S^Q} \,|\, \hat{H}_{\text{SOC}} \, |\, \mathrm{D}, i \, \rangle \right|
\label{eq:soc_mean}
\end{equation}
where $\mathrm{M_S^Q} \in \{\pm 0.5,\, \pm 1.5\}$ is the spin projection of the \textsuperscript{4}T\textsubscript{2g} quartet, $i$ indexes the four doublet components (the M\textsubscript{S} = $\pm 0.5$ projections of each of the two nearly degenerate \textsuperscript{2}E\textsubscript{g} doublets), and $|\cdot|$ denotes the modulus of the complex Breit-Pauli SOC matrix element\cite{fedorov2003spin} as implemented in TeraChem. The mean SOC is essentially time independent over the simulation, around 60--80\,cm$^{-1}$, with only moderate variations between trajectories. This insensitivity of SOC to nuclear motion, together with its appreciable magnitude, suggests that the primary mechanism of coupling between vibrational motion and the \textsuperscript{4}T\textsubscript{2g} $\rightarrow$ \textsuperscript{2}E\textsubscript{g} ISC rate is the modulation of the energy gap, rather than modulation of the SOC.

\subsection{3.3 Characterizing the LMCT State}

\begin{figure}
  \centering
  \includegraphics[width=0.8\linewidth]{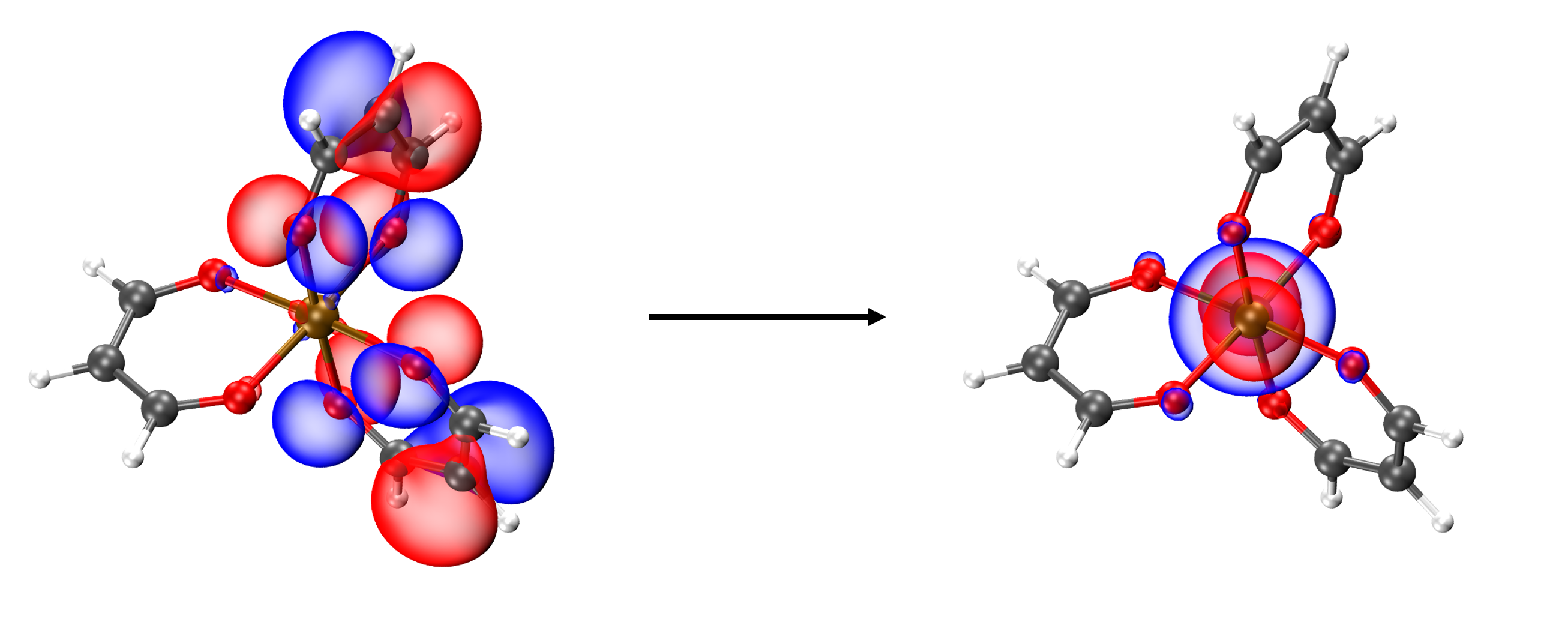}
  \caption{Donor (left) and acceptor (right) orbitals of the \textsuperscript{2}E\textsubscript{g} $\rightarrow$ \textsuperscript{2}LMCT excitation at the \textsuperscript{4}A\textsubscript{2g} optimized geometry. The donor orbital has dominant ligand $\pi$ character, while the acceptor orbital is a metal-centered $t_{2g}$ orbital}
  \label{fig:LMCT}
\end{figure}

Here, we report two representative single-point, vertical excitation calculations aimed at probing the origin of coherent oscillations observed in the pump-probe signals. Note that the prominent ESA feature observed in Ref.~\citenum{schrauben2010vibrational} has been assigned to the \textsuperscript{2}E\textsubscript{g} $\rightarrow$ \textsuperscript{2}LMCT excitation following ISC. We compute XDW-CASPT2 vertical excitation energies for the \textsuperscript{2}E\textsubscript{g} $\rightarrow$ \textsuperscript{2}LMCT transition at two geometries: the \textsuperscript{4}A\textsubscript{2g} minimum and the \textsuperscript{4}T\textsubscript{2g} minimum. (The \textsuperscript{2}E\textsubscript{g} minimum is expected to closely resemble the \textsuperscript{4}A\textsubscript{2g} minimum owing to its intraconfigurational spin-flip character.) The ligand-to-metal character of this transition is illustrated in Figure~\ref{fig:LMCT}, which shows the donor orbital (centered on the ligand) and the acceptor orbital (a Cr d orbital). As summarized in Table~\ref{tab:ESA at two geometries}, the \textsuperscript{2}E\textsubscript{g} $\rightarrow$ \textsuperscript{2}LMCT excitation energy at the \textsuperscript{4}T\textsubscript{2g} minimum (1.67\,eV) is 0.16\,eV smaller than that at the \textsuperscript{4}A\textsubscript{2g} minimum (1.83\,eV). The experimental probe wavelength of 592\,nm (2.09\,eV)\cite{paulus2022use, schrauben2010vibrational} is in reasonable proximity to both values. This analysis supports an interpretation in which coherent oscillations in the TA signal originate from nuclear motion on the \textsuperscript{2}E\textsubscript{g} surface along the reaction coordinate, which modulates the \textsuperscript{2}E\textsubscript{g} $\rightarrow$ \textsuperscript{2}LMCT excitation energy probed in the experiment.

\begin{table}[h!]
  \centering
  \begin{tabular}{ccc}
  \hline
     & \textbf{\textsuperscript{4}A\textsubscript{2g}} & \textbf{\textsuperscript{4}T\textsubscript{2g}} \\ \hline
    ESA/eV & 1.83 & 1.67 \\ \hline
  \end{tabular}
  \caption{XDW-CASPT2 excitation energies of \textsuperscript{2}E\textsubscript{g} $\rightarrow$ \textsuperscript{2}LMCT transition at two representative geometries: \textsuperscript{4}A\textsubscript{2g} and \textsuperscript{4}T\textsubscript{2g} minima.}
  \label{tab:ESA at two geometries}
\end{table}

\section{4. Conclusions}

In summary, BO-AIMD on the excited-state PES \textsuperscript{4}T\textsubscript{2g} of our \ch{Cr(PDO)3} model, combined with electronic structure analysis, elucidates the sub-100\,fs ISC mechanism in photoexcited Cr(III) complexes. The trajectories reveal a persistent JT distortion in the \textsuperscript{4}T\textsubscript{2g} state and a coherent symmetric Cr--O stretching motion, as well, with frequencies of 219\,cm$^{-1}$ and 465\,cm$^{-1}$. The symmetric mode at 219\,cm$^{-1}$ strongly modulates the \textsuperscript{4}T\textsubscript{2g}/\textsuperscript{2}E\textsubscript{g} energy gap, periodically bringing these states into degeneracy, while the SOC remains effectively constant. SA-4-CAS(3,5)SCF calculations of ground state vibrational frequencies identify two modes in good agreement with the FT analysis: 225\,cm$^{-1}$ (twisting/scissoring) and 487\,cm$^{-1}$ (collinear Cr--O stretching). The lower-frequency mode is most likely the ISC-promoting reaction coordinate, consistent with the experimentally observed slowing of ISC upon \textit{tert}-butyl substitution and with prior evidence that twisting motion couples to ISC in other transition metal systems. Finally, the vertical excitation energies for the \textsuperscript{2}E\textsubscript{g} $\rightarrow$ \textsuperscript{2}LMCT transition evaluated at two representative geometries confirm the electronic origin of the experimental probe used to track excited-state coherence. Collectively, these theoretical results provide a mechanistic foundation for predictive tuning of excited-state lifetimes and other properties in transition metal complexes through structure--property relationships.

\begin{acknowledgement}
This work was supported by U.S. Department of Energy, Office of Science, Office of Basic Energy Sciences, under Award No. DE-SC0021643 and by the Institute for Advanced Computational Science at Stony Brook University. JKM was supported by the National Science Foundation under grant CHE-2154233.  We acknowledge Dr. Lixin Lu for assistance with the spin--orbit coupling codes in TeraChem.
\end{acknowledgement}
\begin{suppinfo}
The supporting information document includes optimized geometries and absolute energies for the lowest three electronic states studied in this work.
\end{suppinfo}

\bibliography{Cr(acac)3}

\end{document}